\documentstyle[12pt]{article}
\newcommand{\beq}{\begin{equation}}
\newcommand{\eeq}{\end{equation}}
\newcommand{\bea}{\begin{eqnarray}}
\newcommand{\eea}{\end{eqnarray}}

\newcommand{\k}{\kappa}
\newcommand{\e}{\epsilon}

\newcommand{\rp}{\sqrt{-i \epsilon \hbar}}
\newcommand{\rn}{\sqrt{i \epsilon \hbar}}
\newcommand{\rg}{\sqrt{g}}
\newcommand{\pp}{\pi_\psi}
\newcommand{\p}{\phi}

\newcommand{\tp}{\tilde{\psi}}
\renewcommand{\d}{\delta}

\renewcommand{\b}{\beta}
\renewcommand{\a}{\alpha}
\newcommand{\E}{{\cal E}}
\newcommand{\ER}{\sqrt{\cal E}}

\newcommand{\A}{\mbox{\AE}}
\newcommand{\G}{{\cal G}}
\renewcommand{\H}{{\cal H}}

\newcommand{\N}{\tilde{N}}

\newcommand{\n}{\nu}
\newcommand{\m}{\mu}

\newcommand{\s}{\sigma}
\newcommand{\D}{\Delta}
\newcommand{\th}{\theta}

\newcommand{\ops}{\overline{\psi}}
\newcommand{\oh}{\frac{1}{2}}
\newcommand{\oq}{\frac{1}{4}}

\newcommand{\gb}{\overline{g}}
\newcommand{\rE}{\sqrt{\E}}
\newcommand{\non}{\nonumber}

\renewcommand{\t}{\tau}
\newcommand{\rf}[1]{(\ref{#1})}
\newcommand{\ra}{\rightarrow}

\newcommand{\pa}{\partial}

\begin{document}

\addtolength{\baselineskip}{0.20\baselineskip}

\hfill NORDITA-94/71 P

\hfill gr-qc/9502023

\begin{center}

\vspace{24pt}

{ {\Large \bf Square-Root Actions, Metric Signature, and the
Path-Integral of Quantum Gravity} }

\end{center}

\vspace{6pt}

\begin{center}
\bea
\non \\
   \begin{array}{l}
      \mbox{\sl A. Carlini}^{~\diamondsuit}
{}~~ \mbox{\sl and J. Greensite}^{~\ast}
      \\
      \\
   ^{\diamondsuit}\mbox{NORDITA, Blegdamsvej 17} \\
   ~~\mbox{DK-2100 Copenhagen \O , Denmark} \\
   ~~\mbox{Email: carlini@nbivax.nbi.dk}
        \\
        \\
   ^{\ast}\mbox{Physics and Astronomy Dept.} \\
   ~~\mbox{San Francisco State University} \\
   ~~\mbox{San Francisco, CA 94117, USA} \\
   ~~\mbox{Email: greensit@stars.sfsu.edu} \\
 \non \end{array}
\non
\eea

\vspace{24pt}

{\bf Abstract}

\end{center}

\bigskip

    We consider quantization of the Baierlein-Sharp-Wheeler form of the
gravitational action, in which the lapse function is determined from the
Hamiltonian constraint.  This action has a square root form, analogous to the
actions of the relativistic particle and Nambu string.  We argue that
path-integral quantization of the gravitational action should be based on
a path integrand $\exp[ \sqrt{i} S ]$ rather than the familiar Feynman
expression $\exp[ i S ]$, and that unitarity requires integration over
manifolds of both Euclidean and Lorentzian signature.  We discuss the
relation of this path integral to our previous considerations regarding
the problem of time, and extend our approach to include fermions.

\vfill

\newpage

\section{Introduction}

    Square-root Lagrangians are a feature of many field theories which
are invariant under a time-reparametrization.  The action
of a relativistic particle
\beq
        S_P = -m \int d\t \; \sqrt{-g_{\m\n} \pa_\t x^\m  \pa_\t x^\n}
\label{rp}
\eeq
and the action of the Nambu string
\beq
        S_N = T \int d\s d\t \;  \sqrt{ (\pa_\t \vec{x} )^2
(\pa_\s \vec{x} )^2   - (\pa_\t \vec{x} \cdot \pa_\s \vec{x})^2  }
\label{Nambu}
\eeq
are familiar examples.  Somewhat less familiar is the Baierlein-Sharp-Wheeler
(BSW) form of the gravitational action
\beq
S_{BSW} =-{1 \over \k^2} \int d^4x \;
\sqrt{\sqrt{g} {~}^3R G^{ijnm}(\pa_t g_{ij} - 2N_{(i;j)})
(\pa_t g_{nm} - 2N_{(n;m)}) }
\label{BSW}
\eeq
which is obtained from the standard ADM action, as reviewed below,
by solving the Hamiltonian constraint.  It is well known that for
a relativistic particle moving in an arbitrary curved background, and for
gravity in general, the corresponding quantum theory lacks
a well-defined probability measure and time-evolution parameter
\cite{Kuchar1}, \cite{Kuchar}, \cite{Isham}.

    In this article we will propose a path-integral formulation
of these "square-root" theories which is something of a departure
from the standard Feynman expression.  For one thing, the
integrand of our path-integral will involve an unconventional
phase:
\beq
          \exp[\sqrt{i} S] ~~~~\mbox{rather than}~~~~\exp[iS]
\eeq
Secondly, we will regularize the integration measure so as to
uncover what we believe to be the true time-evolution parameter
of the quantum theory.   Third, we will find it necessary to sum
over path-segments of both real and imaginary proper-time, i.e.
over time-like {\it and} space-like trajectories
in the case of the relativistic
particle; Lorentzian {\it and} Euclidean signature manifolds in the
case of gravity.  It will be shown
that the combination of the regularization, the unconventional
phase, and the inclusion of imaginary proper-time segments,
leads to a unitary evolution of states which corresponds,
via the Ehrenfest principle, to the standard classical dynamics.

    In two previous articles \cite{Me}, \cite{Us}  we have advocated
a transfer-matrix approach to quantizing time reparametrization-invariant
theories.  The present article essentially presents the "real-time"
version of our former "Euclidean" approach. Our previous work did
not include fermion fields, which involve certain complications
in our formulation.  In this paper, we will show how the fermionic
fields are also incorporated into our approach.

\section{Minisuperspace Actions}

    We begin by considering simple quantum-mechanical theories with
a time-reparametrization invariance, i.e. the "minisuperspace"
models of the form
\bea
      S &=& \int dt \; \{ p_a \partial_t q^a  - NH[p,q] \}
\non \\
      H &=& {1 \over 2m} G^{ab}(q) p_a p_b + m V(q)
\label{SH}
\eea
where it is assumed that the supermetric $G^{ab}$ has Lorentzian
signature $(-+++...+)$.
The "square-root" form of the action is obtained by solving
for $p_a$ in terms of the time-derivatives of the $\{q^a\}$,
i.e.
\bea
    & & \partial_t q^a = N{\partial H \over \pa p_a} = {NG^{ab}p_b\over m}
\non \\
    &\Longrightarrow& p_a = {m \over N} G_{ab} \partial_t q^b
\label{find_p}
\eea
and then solving the Hamiltonian constraint for the lapse function
\bea
      0 &=& {1 \over 2m} G^{ab}(q) p_a p_b +  m V(q)
\non \\
        &=& {m\over 2N^2} G_{ab} \partial_t q^a \partial_t q^b
                + m V(q)
\non \\
  \Longrightarrow N &=& \sqrt{ -{G_{ab} \over 2 V}
\partial_t q^a \partial_t q^b }
\label{find_N}
\eea
Substituting \rf{find_p} and \rf{find_N} into the minisuperspace
action then gives the square-root form
\beq
     S = - m \int dt \; \sqrt{-2V G_{ab} \partial_t q^a \partial_t q^b }
\label{sr}
\eeq
For $V=\oh$, this is simply the action for a relativistic particle
of mass $m$, moving in a background manifold with metric $G_{ab}$.

   In non-relativistic quantum mechanics, a path-integral is
constructed out of elementary integrals which evolve the
wavefunction by a small time-interval $\e$, i.e.
\bea
     \psi(x',t+\e) &=& \int d^Dx \; \m_\e \exp[iS[(x',t+\e);(x,t)]/\hbar]
\psi(x,t)
\non \\
                 &=& U_\e \psi(x',t)
\label{step}
\eea
where $S[(x',t+\e);(x,t)]$ is the action of a classical trajectory
between the points $x$ at time $t$ and $x'$ at time $t+\e$.  The
measure $\m_\e$ is chosen so that $\psi(x,t+\e) \ra \psi(x,t)$
as $\e \ra 0$.   With this rule, one finds
\beq
       U_\e = \exp[-iH\e/\hbar]~~+~~O(\e^2)
\label{Ue}
\eeq
where $H$ (the Hamiltonian) is an
$\e$-independent Hermitian operator. Taking the $\e \ra 0$ limit,
the evolution operator for finite times
\bea
      U_{\D t} &\equiv& \lim_{\e \ra 0} (U_\e)^{\D t/\e}
\non \\
               &=& \exp[-iH\D t/\hbar]
\eea
is a unitary operator. Straightforward
imitation of this construction doesn't work in the case of the
square-root theories,  due to the time-reparametrization invariance.
Because of this invariance, the action of a classical trajectory
between an initial point $q$ and an
end point $q'$ is
independent of the time parameters $t$ and $t+\e$ which label those
configurations, i.e.
\beq
         S[(q',t+\e);(q,t)] = S[q',q]
\eeq
The resulting operator $U_\e$ defined from \rf{step} would therefore
be $\e$-independent, and also in general non-unitary.

     Let us see if it is possible to recover an evolution operator
of the form \rf{Ue} for the square-root actions, by making a slight change
to the construction shown in eq. \rf{step}.  The modification
is to multiply the action $S[q',q]$ by an $\e$-dependent
complex constant $c_\e$
\bea
     \psi(q',\t+\e) &=& \int d^Dq \; \m_\e \exp[c_\e S(q',q)]
                       \psi(q,\t)
\non \\
                   &=& U_\e \psi(q',\t)
\label{modify}
\eea
which is to be chosen such that $U_\e$ is a unitary operator
(up to order $\e$) of the form
\beq
       U_\e = \exp[-i\A \e /\hbar]~~+~~O(\e^2)
\label{Ue1}
\eeq
where $\A$ is an $\e$-independent operator, hermitian in the measure
$\m_\e$.  Begin, for simplicity, with a "minisuperspace" action
having $V=\oh$ and $G_{ab}=\eta_{ab}$; i.e. the action of a relativistic
particle in flat D-dimensional Minkowski space.  Let
\beq
        x'^\m = x^\m - \D x^\m
\eeq
so that
\beq
     \psi(x',\t+\e) = \int d^Dx \; \m_\e \exp[-c_\e m
     \sqrt{-\eta_{ab} \D x^a \D x^b }] \psi(x,\t)
\label{step1}
\eeq
Comparing this expression to the corresponding expression for
a free non-relativistic particle
\beq
     \psi(x',\t+\e) = \int d^Dx \; \m_\e \exp\left[ m {
     \d_{ij} \D x^i \D x^j \over (-i \e \hbar) }\right] \psi(x,t)
\eeq
motivates us to try
\beq
        c_\e = {1 \over \sqrt{-i \e \hbar} }
\label{ce}
\eeq
with the understanding that the "time"-step $\e$ now has units of
action, and that the branch of the square-root is chosen so that
the exponent in  \rf{step1} has a negative real part.  This choice
does not quite complete the definition of $U_\e$, as there is still a
question of the range of the integral over $x$.  Should this integral
range over all possible $x$, or should it be restricted so that
the path-segment $\D x^a$ is timelike?  To resolve this issue,
we will compute separately the contributions from timelike and spacelike
intervals.

    Following the usual steps leading from the path integral to the
Schr\"odinger equation, expand $\psi(x,\t)$ in a Taylor series around
$x'$
\bea
    \psi(x',\t+\e) &=& \int d^Dx \; \m_\e \exp[- m
     \sqrt{-\eta_{\m \n} \D x^\n \D x^\m }/\rp ]
\non \\
  & & \times \left[ \psi(x',\t) +
 {\partial \psi \over \partial x'^\m} \D x^\n
  + \oh {\partial^2 \psi \over \partial x'^\m \partial x'^\n} \D x^\m \D x^\n
+ ... \right]
\non \\
  &=& U_\e \psi(x',t)
\eea
In order that
\beq
      \lim_{\e \ra 0} U_\e = 1
\eeq
the measure must be chosen to be
\beq
      \m_\e^{-1} =  \int d^Dx \; \exp[-m\sqrt{-\eta_{\m \n} \D x^\m
                       \D x^\n }/\rp ]
\eeq
Changing variables $\D x \ra x$, we then have
\beq
 U_\e = 1 + \oh \left[ \int d^Dx \; \m x^\m x^\n
\exp[-m\sqrt{-\eta_{\m\n} x^\m x^\n}/\rp] \right] \pa_\m \pa_\n
{}~~+~~...
\eeq
Denote $x^\m = \{t,\vec{x}\}$ and $r^2 = \vec{x} \cdot \vec{x}$.
Then, on grounds of relativistic covariance,
\bea
 U_\e &=& 1 + \oh \left[ \int d^Dx \; \m {r^2 \eta^{\m\n} \over (D-1)}
\exp[-m\sqrt{-\eta_{\m\n}x^\m x^\n}/\rp] \right] \pa_\m \pa_\n
{}~~+~~...
\non \\
   &=& 1 + {1 \over 2(D-1)} {I_B \over I_A} \pa^2 ~~+~~...
\label{Ue2}
\eea
where $\pa^2=\eta^{\m \n}\pa_\m \pa_\n$, and
\bea
     I_A &\equiv& \int d^Dx \; \exp[-m\sqrt{-\eta_{\m\n}x^\m x^\n}/\rp]
\non \\
     I_B &\equiv& \int d^Dx \; r^2 \exp[-m\sqrt{-\eta_{\m\n}x^\m x^\n}/\rp]
\eea

   We first evalute $I_A$; the second integral $I_B$ will
follow easily.  Starting with
\beq
     I_A = \s \int dt dr \; r^{D-2} \exp[-m\sqrt{t^2-r^2}/\rp]
\eeq
where
\beq
      \s = {2\pi^{(D-1)/2} \over \Gamma({D-1 \over 2}) }
\eeq
divide the integral over $t$ into two contributions, one from
timelike and one from spacelike paths:
\bea
      I_A &=& \s \int_0^\infty dr r^{D-2} \left\{ \int_r^{\infty} dt
\exp[-m\sqrt{t^2-r^2}/\rp]\right.
\non \\
& & \left. + \int_0^r dt \exp[-m\sqrt{r^2-t^2}/\rn]\right\}
\eea
where the branches of the square roots $\rp$ and $\rn$ are taken with
positive real parts, to ensure convergence of the integrals.
Next
\bea
  I_A &=& \s \int_0^\infty dr r^{D-2} \left\{ \int_r^{\infty} dt
\exp[-m\sqrt{t^2-r^2}/\rp]\int_0^\infty dy 2y \d[y^2-(t^2-r^2)] \right.
\non \\
 & & \left. + \int_0^r dt \exp[-m\sqrt{r^2-t^2}/\rn]\int_0^\infty dy
\; 2y \d[y^2 - (r^2-t^2)] \right\}
\non \\
 &=& \s \int_0^\infty dr r^{D-2} \int_0^\infty dy \left\{
{y \over \sqrt{y^2+r^2}} e^{-my/\rp} +
{y \over \sqrt{r^2-y^2}} e^{-my/\rn} \Theta(r-y) \right\}
\non \\
 &=& \s \int_0^\infty dr r^{D-2} \{F_1(r) + F_2(r)\}
\label{IandF}
\eea
where
\bea
  F_1(r) &\equiv& \int_0^\infty dy {y \over \sqrt{y^2+r^2}} e^{-my/\rp}
\non \\
  F_2(r) &\equiv& \int_0^r dy {y \over \sqrt{r^2-y^2}} e^{-my/\rn}
\label{F12}
\eea

   It is easy to see that asymptotically, as $r \ra \infty$, both $F_1(r)$
and $F_2(r)$ go like $1/r$.  But that implies
\bea
  \mbox{timelike paths contribution} = \s \int_0^\infty dr r^{D-2} F_1(r)
{}~~~~~\mbox{is divergent}
\non \\
  \mbox{spacelike paths contribution} = \s \int_0^\infty dr r^{D-2} F_2(r)
{}~~~~~\mbox{is divergent}
\eea
This means that if we were to restrict the paths to only timelike,
or only spacelike paths, then $I_A$ (and $I_B$) would be hopelessly
divergent, and the evolution operator $U_\e$ would be ill-defined.
The remarkable thing, which we now show, is that the sum of
the two contributions is actually finite.

   Let us deform the contour of $y$-integration for the integral defining
$F_1(r)$ in eq. \rf{F12}.  As it stands, it runs along the real axis
from $0$ to $\infty$.  Deform it to run along the imaginary axis from
$0$ to $-ir$, and then parallel to the real axis from $-ir$ to $\infty$.
There are no poles or branch cuts in the way, so the deformation
is permissible.  Then

\beq
     F_1(r) = \left\{ \int_0^{-ir} dy + \int_{-ir}^\infty dy \right\}
{y \over \sqrt{r^2 + y^2}} e^{-my/\rp}
\eeq
Change variables, $y \ra -iy$
\beq
     F_1(r) = -\int_0^r dy {y \over \sqrt{r^2 - y^2}} e^{-my/\rn}
- \int_{r}^{i\infty} dy {y \over \sqrt{r^2 - y^2}} e^{-my/\rn}
\label{F3}
\eeq
Its not hard to see that it is the first integral which causes the
divergence of the $r$-integration.  Adding together $F_1$ and $F_2$,
observe that the first integral in \rf{F3} exactly cancels $F_2$,
leaving an expression which decays exponentially as $r$ increases
\beq
     F_1+F_2 = - \int_r^{i\infty} dy {y \over \sqrt{r^2 - y^2}} e^{-my/\rn}
\eeq
The contour of this integral runs parallel to the (positive) imaginary
axis.  Now rotate the contour by 90 degrees, so that it runs along the
positive real axis.  Again, there are no poles or branch
cuts in the way, and the integral is convergent along any contour
intermediate between the initial contour, and the 90 degree rotated
contour.  This gives
\bea
     F_1+F_2 &=& i \int_r^\infty dy {y \over \sqrt{y^2-r^2}} e^{-my/\rn}
\non \\
     &=& ir K_1({mr \over \rn})
\eea
Inserting this result into \rf{IandF}, one finds
\bea
     I_A &=& i \s \int_0^\infty dr r^{D-1} K_1({mr \over \rn})
\non \\
         &=& i \s 2^{D-2} ({\rn \over m})^{D} \Gamma({D+1 \over 2})
\Gamma({D-1 \over 2})
\label{IA}
\eea
It is trivial to repeat all the above steps for the $I_B$ integral,
and the result is
\bea
     I_B &=& i \s \int_0^\infty dr r^{D+1} K_1({mr \over \rn})
\non \\
         &=& i \s 2^D ({\rn \over m})^{D+2} \Gamma({D+1 \over 2}+1)
\Gamma({D-1 \over 2}+1)
\label{IB}
\eea
Finally, inserting \rf{IA} and \rf{IB} into the expression for the
evolution operator, eq. \rf{Ue2}, we obtain
\bea
   U_\e &=& 1 + i\e \hbar {(D+1) \over 2m^2} \eta^{\m\n} \pa_\m \pa_\n +
O(\e^2)
\non \\
         &=& \exp[-i\A\e/\hbar] ~~+~~O(\e^2)
\label{Ue3}
\eea
where
\beq
      \A = -{\hbar^2 \over 2m^2}(D+1)\eta^{\m \n} \pa_\m \pa_\n
\label{relpart}
\eeq
The operator $\A$ is $\e$-independent, and clearly Hermitian
for inner products
\beq
      <\psi_1|Q|\psi_2> = \int d^Dx \; \m_\e \psi_1^*(x,\t) Q
                  \psi_2(x,\t)
\eeq
The evolution operator
\bea
      U_{\D \t} &\equiv& \lim_{\e \ra 0} (U_\e)^{\D \t /\e}
\non \\
               &=& \exp[-i\A \D \t /\hbar]
\eea
is therefore unitary, as
in the usual path-integral approach for theories without
a time-reparametrization invariance.

   It should be emphasized that the unitarity of our proposed
evolution operator depends both on the choice of complex constant
$c_\e = 1/\sqrt{-i\e \hbar}$, and also on summation over both
timelike and spacelike path segments.  A glance at equations \rf{IA}
and \rf{IB} shows that the crucial factor of $i\e$ in \rf{Ue3}
could only be obtained if a $1/\sqrt{-i\e}$ factor multiplies
the action in \rf{modify}.  Furthermore, the finiteness of the
result depends on keeping contributions to the integrand from both
timelike and spacelike path-segments; the integral over either contribution
separately is divergent.

   It is easy to generalize from the relativistic particle action
to any minisuperspace square-root action of the form \rf{sr}.
First define the modified supermetric
\beq
      \G_{ab} \equiv 2VG_{ab}
\eeq
For $\D q^a$ small
\bea
      S[q',q] &=& - m \int_0^{\D t} dt \; \sqrt{-\G_{ab} \pa_t q^a
\pa_t q^b}
\non \\
              &=& -m \sqrt{-\G_{ab} \D q^a \D q^b}
\eea
The measure is
\beq
\m_\e^{-1}(q') = (\sqrt{\e \hbar})^D \lim_{\e \ra 0}
\int {d^D q \over (\sqrt{\e \hbar})^D }
 \; \exp[-m\sqrt{-\G_{ab} \D q^a \D q^b}/\sqrt{-i\epsilon \hbar}]
\eeq
Now introduce Riemann normal coordinates $\xi^a$ around the point
$q'^a$, which transforms the modified supermetric into
the Minkowski metric $\G_{ab}=\eta_{ab}$ at the point $q'$ ($\xi=0$).
Then
\bea
    \m_\e^{-1}(q') &=& \int d^D \xi \; \left| \det[{\pa \D q^a \over
\pa \xi^b}] \right| \exp[-m \sqrt{-\eta_{\a \b} \xi^a \xi^b}/\rp]
\non \\
                  &=& {1 \over \sqrt{|\G(q')|} } I_A
\label{me}
\eea
Inserting this measure, and $c_\e = 1/\sqrt{-i\e \hbar}$, into
eq. \rf{modify}, we have
\bea
       \psi(q',\t+\e) &=& {1 \over I_A} \int d^D \D q
 \sqrt{\vert\G(q'+ \D q)\vert}  \exp[-m\sqrt{-\G_{ab}\D q^a \D q^b}/\rp ]
\non \\
& & \times \psi(q'+\D q,\t)
\non \\
   &=&  {1 \over I_A}\int d^D\xi \;
(1 - {1\over 6}{\cal R}_{a b} \xi^a \xi^b + ...)
 \exp[-m \sqrt{-\eta_{a b}\xi^a \xi^b} / \rp]
\non \\
& & \times \left\{ \psi(q',\t) +
{\partial \psi \over \partial \xi^c}\xi^c
 + \oh{\partial^2 \psi \over \partial \xi^c \partial \xi^d}
\xi^c \xi^d + O(\xi^3) \right\}
\non \\
   &=& \left[ 1 + {1 \over (D-1)}{I_B \over I_A} \{ \oh \partial^2 -
 {1 \over 6} {\cal R} + O(\e^2)\} \right] \psi(q'(\xi),\t)
\non \\
   &=& \left[ 1 + i\e \hbar {(D+1) \over 2m^2}\eta^{ab} {\partial^2 \over
\pa \xi^a \pa \xi^b} - i\e \hbar {(D+1) \over 6m^2}{\cal R} \right]
\psi(\xi,\t)
\eea
where ${\cal R}$ is the curvature scalar formed from the metric $\G_{ab}$.
Transforming back from Riemann
normal coordinates, we have
\bea
  \psi(q,\t+\e) &=& \left[ 1 + i\e \hbar {(D+1) \over 2m^2}
{1 \over \sqrt{|\G|}} {\partial \over \partial q^a} \sqrt{|\G|} \G^{ab}
{\partial \over \partial q^b} - i\e \hbar {(D+1) \over 6m^2}
      {\cal R} \right] \psi(q,\t)
\non \\
         &=& \left[ \exp[-i\A \e/\hbar] + O(\e^2) \right] \psi(q,\t)
\non \\
         &=& U_\e \psi(q,\t)
\eea
As in the relativistic particle case, $\exp[-i\A \e/\hbar]$ is
a unitary operator, where
\beq
      \A = -  \hbar^2  {(D+1) \over 2m^2} {1 \over \sqrt{|\G|}} {\partial
\over \partial q^a} \sqrt{|\G|} \G^{ab} {\partial \over \partial q^b}
 +  \hbar^2 {(D+1) \over 6m^2} {\cal R}
\label{AE}
\eeq
is obviously Hermitian in the measure $\m_\e$ of eq. \rf{me}.  Taking
the $\e \ra 0$ limit, the wavefunction $\psi(q,\t)$ satisfies a
Schr\"odinger equation
\beq
       i \hbar \pa_\t \psi(q,\t) = \A \psi(q,\t)
\label{Seq}
\eeq

   The Schr\"odinger
evolution equations \rf{AE} and \rf{Seq} for time
reparametrization invariant theories have been obtained by us previously,
in refs. \cite{Me} and \cite{Us}, from a transfer matrix approach.
The transfer matrix involves integration over a purely real integrand,
but in our case the cost was not simply a Wick rotation of the evolution
parameter $\t$, but also a rather unnatural rotation of signature of
the modified supermetric $\G_{ab}$ from Lorentzian to Euclidean.
This rotation then had to be undone in postulating the Schr\"odinger
equation \rf{Seq}.  We have now seen that this supermetric signature
rotation can be avoided, and unitary Schr\"odinger evolution is derived
directly.   In refs. \cite{Me} and \cite{Us} the
correspondence of this evolution to the usual classical dynamics was also
discussed.  In the interest of completeness we will briefly
review this correspondence here, and
refer the reader to the cited references for further details.

    The classical quantity $\A_{cl}$ corresponding to the operator $\A$
is obtained by replacing derivatives with c-number momenta
\beq
     \A_{cl}[q^a,p_a] = \lim_{\hbar \ra 0}
     \A[q^a,-i\hbar{\partial \over \partial q^a} \ra p_a]
\eeq
which gives
\beq
      \A_{cl} = (D+1){ {1 \over 2m} G^{ab} p_a p_b \over m V}
\eeq
The Poisson bracket evolution equation
\bea
    \pa_\t Q = \{Q,\A_{cl} \} ~~,~~ \A_{cl} = -\E
\label{PB1}
\eea
is easily checked to be equivalent, up to a time-reparametrization,
to the standard brackets
\bea
    \pa_t Q = \{Q,N\H^\E \} ~~,~~ \H^\E = 0
\label{PB2}
\eea
where
\beq
      \H^\E = {1 \over 2m \sqrt{\E}} G^{ab}p_a p_b + \sqrt{\E} mV
\label{HE}
\eeq
The parameter $\E$ is classically irrelevant, in the
sense that it drops out of the Euler-Lagrange equations; the
square-root action corresponding to $\H^\E$ is
\beq
  S^\E = - \sqrt{\E} m \int dt \; \sqrt{-2V G_{ab} \pa_t q^a \pa_t q^b }
\eeq
Since $\E$ only appears as a parameter multiplying the action, the fact
that it drops out of the Euler-Lagrange equations is obvious.  The
same can be said for the mass of a relativistic particle in free-fall,
the tension of the Nambu string, or Newton's constant in pure gravity.
None of these parameters appears in the equations of motion at the
classical level.

   Because of the classical equivalence of the Poisson bracket
equations \rf{PB1} and \rf{PB2}, it is clear that the
Schr\"odinger
evolution \rf{Seq} will obey an appropriate Ehrenfest principle,
with certain quantum corrections due to the measure.  The general
solution $\psi(q,\t)$ can be expanded in terms of stationary states
\beq
      \psi(q,\t) = \sum_\E \phi_\E (q) e^{i\E \t/\hbar}
\eeq
where
\bea
           \A \phi_\E = -\E \phi_\E
\eea
This $\t$-independent equation can be rewritten in the form
\bea
   \left[ - {\hbar^2 \over \E}  {(D+1) \over 4m^2}
{V \over \sqrt{|\G|}} {\partial \over \partial q^a} V^{-1} \sqrt{|\G|} G^{ab}
{\partial \over \partial q^b} +  {\hbar^2 \over \E}
{(D+1) \over 6m^2} V{\cal R}
     + V \right] \phi_\E(q) = 0
\label{WD}
\eea
which is a Wheeler-DeWitt equation with a particular choice of
operator-ordering and a (classically irrelevant) parameter $\E$,
which can be absorbed into a redefinition of either $m$ or $\hbar$.

   In the standard Dirac canonical quantization of actions of the
form \rf{SH}, the physical states must satisy a
Wheeler-DeWitt equation of the form
\bea
  \H \phi(q) &=& \left[{-\hbar^2 \over 2m}"G^{ab}{\pa^2 \over \pa q^a \pa q^b}"
           + mV \right] \phi(q)
\non \\
             &=& 0
\eea
where the quotation marks indicate an operator-ordering ambiguity.
However, multiplying the action by an arbitrary
constant $\sqrt{\E}$ converts this constraint to
\bea
  \H^\E \phi(q) &=& \left[{-\hbar^2 \over 2m \sqrt{\E} }
"G^{ab}{\pa^2 \over \pa q^a \pa q^b}" + \sqrt{\E} mV \right] \phi(q)
\non \\
             &=& 0
\eea
Because $\E$ is irrelevant at the classical level, there is no
overriding reason that it should be a fixed parameter at the
quantum level.  In essence, our approach enlarges the space of physical
states as the Hilbert space spanned by all states satisfying
\beq
       \H^\E \phi_\E(q) = 0
\eeq
and it is this enlargement of the space of states which enables us
to obtain non-stationary states $\psi(q,\t)$.  Moreover, our approach
fixes the operator-ordering, as seen in eq. \rf{WD}, at least for
the minisuperspace theories.  Further discussion
of these points may be found in ref. \cite{Us}.

   Returning to eq. \rf{modify}, the path integral for square-root
actions is now defined as the limit
\bea
      \psi(q_f,\t_0+\D \t) &=& \int Dq(\t_0\le \t<\t_0+\D \t) \;
e^{c_0 S[q(\t)]}  \psi(q_0,\t_0)
\non \\
         &=& \lim_{\e \ra 0} \int \; \prod_{n=0}^{\D \t /\e-1}
d^D q_n \m_\e(q_n) \exp\left[ {1 \over \sqrt{-i\e \hbar}}
 \sum_{m=0}^{\D \t /\e-1} S[q_{m+1},q_m] \right] \psi(q_0,\t_0)
\non \\
         &=& \lim_{\e \ra 0} (U_\e)^{\D \t/\e} \psi(q_f,\t_0)
\eea
where $\D \t = \t_f - \t_0$.  We see that the time parameter
emerges from a regularization of the path-integral measure: at
fixed $\e$, a regularized path between the initial point $q_0$
and the final point $q_f$  consists of $n_p=\D \t /\e$ path segments,
each segment being a classical trajectory between intermediate
points $q_n$ and $q_{n+1}$.  The evolution parameter is therefore
a measure of the number of independent configurations (points) $n_p$
in the path joining $q_0$ to $q_f$, multiplied by the regularization interval,
i.e.
\beq
        \D \t = n_p \e
\label{tdef}
\eeq
This is a quantum-mechanical time variable with
no direct connection to, e.g., the
proper time lapse.  Nor is it an "intrinsic" time variable; all
dynamical degrees of freedom are treated on the same footing and none
is singled out as an evolution variable.
Our evolution parameter is here identified as proportional to the number
of "quantum steps"
taken by the system in evolving from the initial to the final
configuration.  In this formulation the Green's functions are transitive,
and the evolution of states is unitary.

\section{The BSW Action}

   As in the minisuperspace case, the square-root form of the full
gravitational action is derived from the first-order ADM action
by solving for the lapse function.  The ADM action for pure gravity
is
\bea
   S &=& \int d^4x \; [p^{ij}\pa_t g_{ij} - N\H - N_i \H^i]
\non \\
   \H &=& \k^2 G_{ijkl} p^{ij} p^{kl} - {1\over \k^2} \rg {~}^3R
\non \\
   \H^i &=& -2p^{ik}_{~~;k}
\non \\
   G_{ijkl} &=& {1 \over 2\rg}(g_{ik}g_{jl}+g_{il}g_{jk}-g_{ij}g_{kl})
\label{ADM}
\eea
where $g_{ij}$ is the metric of a 3-manifold and ${~}^3R$
is the corresponding scalar curvature.  The momentum is related to
the time-derivative of the metric by
\bea
     \pa_t g_{ij} &=& N{\pa \H \over \pa p^{ij} }
\non \\
            &=& 2\k^2 N G_{ijkl}p^{kl} + N_{i;j} + N_{j;i}
\non \\
  \Longrightarrow p^{ij} &=& {1 \over 2\k^2 N}G^{ijkl}
                                   (\pa_t g_{kl} - 2N_{(k;l)})
\label{p_ij}
\eea
Solving the Hamiltonian constraint for the lapse function
\bea
    0 &=& {1 \over 4\k^2 N^2} G^{ijkl}(\pa_t g_{ij} - 2N_{(i;j)})
             (\pa_t g_{kl} - 2N_{(k;l)}) - {1 \over \k^2}\rg {~}^3R
\non \\
   \Longrightarrow N &=& \left[ {1 \over 4\rg {~}^3R} G^{ijkl}
(\pa_t g_{ij} - 2N_{(i;j)}) (\pa_t g_{kl} - 2N_{(k;l)}) \right]^{1/2}
\label{NN}
\eea
and replacing the momenta in \rf{ADM} by the expression \rf{p_ij},
with lapse \rf{NN}, gives the Baierlein-Sharp-Wheeler action \cite{BSW}
\beq
    S_{BSW} =-{1 \over \k^2} \int d^4x \; \sqrt{\sqrt{g} {~}^3R G^{ijnm}
(\pa_t g_{ij} - 2N_{(i;j)}) (\pa_t g_{nm} - 2N_{(n;m)}) }
\label{BSW1}
\eeq
Before quantizing, it is convenient to fix the coordinate system
by choosing shift functions $N_i=0$.  Then the corresponding
supermomentum contraints $\d S/ \d N_i = \H^i=0$ are to be imposed
as operator constraints on the space of physical states.

   It is straightforward to extend the BSW action to include
non-gravitational bosonic fields.  To compress indices somewhat,
we introduce the notation
\bea
        \{a=1-6\} &\leftrightarrow& \{(i,j),~i \le j \}
\non \\
             q^a(x) &\leftrightarrow& g_{ij}(x)
\non \\
             p_a(x) &\leftrightarrow&
\left\{ \begin{array}{rr}
             p^{ij}(x)~~~~~(i=j) \\
           2 p^{ij}(x)~~~~~(i<j) \\
        \end{array} \right.
\non \\
             G_{ab}(x) &\leftrightarrow&  G^{ijnm}(x)
\eea
and the non-gravitational fields are represented by $q^a(x)$ with
indices $a>6$.  It is convenient to rescale all non-gravitational
fields by an appropriate power of $\k$ so that all fields, and
all components of the supermetric, are dimensionless.  The action is
\bea
     S &=& \int d^4x [p_a \pa_t q^a - N\H - N_i \H^i]
\non \\
     \H &=& \k^2 G^{ab}p_a p_b + \rg U
\eea
where
\beq
  \rg U = - {1 \over \k^2}\rg {~}^3R ~+~ \mbox{non-gravitational contributions}
\eeq
Setting the shift functions to zero and repeating the above steps of
solving for the lapse, gives again a square-root action
\beq
      S = -{1\over \k}\int d^4x \; \sqrt{-\sqrt{g} U G_{ab} \pa_t q^a
                                                       \pa_t q^b }
\eeq

    The next step is to construct the evolution operator $U_\e$
for the BSW action in the path-integral approach, following the
procedure of the last section.  The evolution operator is defined by
\bea
 \Psi[q'(x),\t+\e] &=& \int Dq(x) \; \m(q) e^{\D S/\rp } \left[ \Psi[q'(x)]
+ \int d^3x \left({\d \Psi \over \d q^a(x)}\right) \D q^a(x)
\right.
\non \\
    & & \left. + \oh \int d^3x d^3y \left( {\d^2 \Psi \over \d q^a(x)
\d q^b(y)}\right) \D q^a(x) \D q^b(y) + ... \right]
\non \\
    &=& \Psi(q',\t) + [T_0 + T_1 + T_2] + O(\e^2)
\non \\
    &=& U_\e \Psi(q',\t)
\label{Tn}
\eea
where the $T_n$ represent terms with $n$ functional derivatives of
$\Psi$ and one power of $\e$, and
\bea
\D S &=& -{1\over \k}\int d^3x \; \sqrt{-\sqrt{g} U G_{ab} \D q^a \D q^b }
\non \\
   \D q^a &=& q^a - q'^a
\eea
In order to obtain $U_\e$, we need to evaluate
\bea
    <\D q^a(x_1) \D q^b(x_2)> &=& \int D(\D q) \; (\m)_0
\D q^a(x_1) \D q^b(x_2)
\non \\
& & \times \exp\left[-{1 \over \k}\int d^3x (\rg)_0 \sqrt{-(\G_{ab})_0
\D q^a \D q^b}/\rp \right]
\label{singular}
\eea
where $\G_{ab}$ is the modified supermetric
\beq
     \G_{ab} \equiv {1 \over \rg} U G_{ab}
\eeq
and $()_0$ indicates that the quantity in parenthesis is evaluated
at $q=q'$.

   Clearly, $<\D q \D q>$ is a highly singular quantity, and
is only well-defined in the context of a regularization
procedure.  In the absence of a non-perturbative regulator
which preserves the exact diffeomorphism invariance, we work
with a naive lattice regulator in which the continuous degrees of freedom
labeled by $x$ are replaced by a discrete set, labeled by
$n$, associated with regions of volume $v_n$.  We have in mind, e.g.,
a Regge-style discretization of a continuous 4-manifold into
a fixed number $N_p$ of simplices
of varying volume. As $N_p \ra \infty$, the choice of $v_n$ is of
course required
to be irrelevant in computing the evolution operator, as long as the
regions' volumes $v_n \ra 0$ in this limit.  As we will see, this
requirement is not satisfied trivially or automatically. We take the
naive lattice-continuum correspondences to be
\bea
        \D q^a(x) &\leftrightarrow& \D q^a(n)
\non \\
        \int d^3x \rg &\leftrightarrow& \sum_{n=1}^{N_p} v_n
\non \\
         {\d \over \d q^a(x)}  &\leftrightarrow&
\left( {\d \over \d q^a(n)} \right)_R \equiv {\sqrt{g(n)} \over v_n}
{\partial \over \partial q^a(n)}
\non \\
          Dq &\leftrightarrow& \prod_n d^Dq(n)
\label{discrete}
\eea
With such a discretization, we have
\bea
<\D q^a(n) \D q^b(m)> &=& \int \prod_j d^Dq(j) \; \m \D q^a(n) \D q^b(m)
\non \\
  &\times& \exp \left[ -{1 \over \k} \sum_k v_k \sqrt{-\G_{ab} \D q^a(k)
\D q^b(k)}/\rp \right]
\label{regint}
\eea

   The supermetric $G_{ab}$ for the discretized degrees of freedom
$\{q^a(n)\}$ still has Lorentzian signature, and we can follow
the steps of the last section in integrating over the $q^a$ at
each $n$.  The result is
\beq
  <\D q^a(n) \D q^b(m)> = i\e \hbar (D+1) \k^2 \G^{ab}
       {1 \over v_n^2} \d_{mn}
\label{corr}
\eeq
and we find for the $T_2$ term
\bea
   T_2 &=&  i\e \hbar {(D+1)\over 2} \k^2
\sum_n \G^{ab} {1 \over v_n^2} {\pa^2 \over \pa q^a(n) \pa q^b(n)} \Psi
\non \\
       &=& i\e \hbar {(D+1)\over 2} \k^2 \sum_n \rg
           {1 \over U} G^{ab}{1\over v_n^2} {\pa^2 \over \pa q^a(n) \pa
q^b(n)} \Psi
\non \\
       &=& i\e \hbar {(D+1)\over 2} \k^2 \sum_n
   \left\{ {1 \over \rg U} G^{ab} \left( {\d \over \d q^a(n)} \right)_R
                   \left( {\d \over \d q^b(n)} \right)_R \Psi \right\}
\eea
The term in braces has a simple continuum limit and, if this term
were weighted by a volume factor $v_n$, then the continuum limit
would be simply
\beq
      T_2 = i\e \hbar {(D+1)\over 2} \k^2 \int d^3x \; U^{-1}
           G^{ab} {\d^2 \over \d q^a \d q^b} \Psi
           ~~~~~~ \mbox{(wrong)}
\eeq
There is, however, no such $v_n$ weighting factor in the sum,
which means that the contribution of each term at each position $n$
\beq
{1 \over \rg U} G^{ab} \left( {\d \over \d q^a(n)} \right)_R
                   \left( {\d \over \d q^b(n)} \right)_R \Psi
\label{term}
\eeq
is weighted equally, regardless of the cell or simplex volume $v_n$.
As a consequence, even in the $v_n \ra 0$ limit, the final answer
for the state evolution would seem to depend on the distribution
of volumes $\{ v_n \}$.

   Such regularization dependence never arises in ordinary quantum
field theory.  There may be other regularization issues, such as
renormalization and anomalies, but certainly one doesn't encounter
{\it this} kind of dependence on the distribution of cell volumes in
computing the naive continuum limit of the Hamiltonian operator.
Since the problem doesn't arise in ordinary quantum
field theory, why does it come up here?  The reason, of course,
can be traced back to the square-root form of the action, which gives
a factor of $1/v_n^2$, rather than a factor of $1/v_n$, in the correlator
\rf{corr}.  The additional power of $1/v_n$ is the source of the
(apparent) difficulty.  There is only one way out, if the evolution operator
is not to depend on the $\{ v_n \}$ distribution: {\it we must
impose a constraint on the physical states $\Psi$ such that
the term \rf{term} is
independent of the discretized position label $n$}, at least in
the $v_n \ra 0$ limit.  In that case, we have
\beq
       T_2 = i \e \hbar {(D+1) N_p \over 2} {1 \over \rg U}
          \k^2 G^{ab} \left ({\d \over \d q^a(n)}\right )_R\left (
{\d \over \d q^a(n)}\right )_R \Psi
              ~~~~~~ \mbox{(any $n$)}
\eeq
The other terms  $T_0$ and $T_1$ are operator-ordering contributions which,
in the absence of an exact diffeomorphism-invariant regulator, will not
be considered further here.  Now absorbing the factor $\oh (D+1) N_p$
into a redefinition of $\e$, and taking the continuum limit, we
arrive at
\beq
       i\hbar {1\over \rg U} \k ^2 G^{ab} {\d^2 \over \d q^a \d q^b} \Psi
                  = \pa_\t \Psi  ~~~~~~ \mbox{(all $x$)}
\label{evolve}
\eeq
Expanding
\beq
       \Psi(q,\t) = \sum_{\E} a_\E e^{i\E \t /\hbar} \Phi_\E(q)
\eeq
eq. \rf{evolve} requires that for each stationary state
\beq
{1\over \sqrt{\E}} \H^\E \Phi_\E =
\left\{ -\hbar^2 {\k^2 \over \E} G^{ab} {\d^2 \over \d q^a \d q^b}
+  \rg U \right\} \Phi_{\E} = 0
\label{WD2}
\eeq
which is simply the Wheeler-DeWitt equation (up to operator-ordering
contributions), with an effective value of Planck's constant
rescaled by
\beq
       \hbar_{eff} = {\hbar \over \sqrt{\E}}
\eeq
Moreover, the Wheeler-DeWitt equation is consistent with, and in
fact implies (via the Moncrief-Teitelboim interconnection theorem
\cite{Moncrief}), the supermomentum constraints
\beq
       \H^i \Phi_\E = 0
\label{Hi_constraint}
\eeq
which are needed to compensate the gauge choice $N_i=0$. Thus, each
stationary state $\Phi_\E$ satisfies the usual constraint algebra
of general relativity, with a rescaled value of Planck's
constant.  The Hilbert space
of all physical states is spanned by the stationary states, with all
possible values of $\E$.  Finally, multiplying both sides of
\rf{evolve} by $N\rg U$, where $N$
is an arbitrary function, integrating over space, and applying
the supermomentum constraint \rf{Hi_constraint}, we obtain the
equation of motion
\bea
   i\hbar \partial_\t \Psi &=& \left[ {1 \over \int d^3x' \rg N U}
 \k^2 \int d^3x \;   N G^{ab} (-\hbar^2{\d^2 \over \d q^a \d q^b})
\right] \Psi
\non \\
   &=& {1 \over m_P} \int d^3x \; \left[ -\hbar^2  \N \k^2 G^{ab}
{\d^2 \over \d q^a \d q^b} + N_i \H^i_x \right] \Psi
\non \\
   &=& \A \Psi
\label{AE2}
\eea
where
\beq
   \A = {1 \over m_P} \int d^3x \; \left[ -\hbar^2  \N \k^2 G^{ab}
{\d^2 \over \d q^a \d q^b} + N_i \H^i_x \right]
\eeq
and
\beq
         \N(x) \equiv  m_P {N(x)  \over \int d^3x' \; \rg N U(q) }
\eeq
with $m_P$ an arbitrary parameter of dimension of mass.

   The evolution equation \rf{AE2} was obtained in ref. \cite{Us} by a
transfer matrix approach, and shown to correspond to the usual
classical evolution via the Ehrenfest principle.  Here we have
instead used the path integral to obtain a unitary evolution operator,
as in the real-time Feynman
approach, and avoided the signature rotation
of the supermetric which was required in deriving the transfer matrix.
The main point of this section is that, in performing the "real-time"
path-integral of the BSW action
\bea
   \lefteqn{ \Psi[q_f(x),\t_0+\D \t] }
\non \\
&=& \int Dq(x,\t) \;
e^{c_0 S[q(\t)]}  \Psi[q_0(x),\t_0]
\non \\
         &=& \lim_{\e \ra 0} \int \; \prod_{n=0}^{\D \t /\e-1}
Dq^a_n(x) \m_\e[q_n] \exp\left[ {1 \over \sqrt{-i \e \hbar}}
 \sum_{m=1}^{\D \t /\e-1} S[q_{m+1}(x),q_m(x)] \right]
\non \\
         & & \times \Psi[q_0(x),\t_0]
\non \\
         &=& \lim_{\e \ra 0} (U_\e)^{\D \t/\e} \Psi(q_f(x),\t_0)
\non \\
         &=& e^{-i\A \D \t /\hbar} \Psi[q_f(x),\t_0]
\eea
it is necessary, as in the minisuperspace case, to integrate over
{\it all} possible paths, including those for which the lapse
function
\beq
N(x) = \left[ -{1 \over 4\k^2 \rg U}G_{ab}\pa_t q^a \pa_t q^b \right]^{1/2}
\eeq
is imaginary.  Real-valued lapse functions correspond to Lorentzian
4-manifolds, imaginary values correspond to Euclidean signature.
If the paths are restricted to real $N(x)$ only,
then we find that due to the Lorentzian signature of the supermetric,
the integrals in eq. \rf{regint} are singular despite the regularization.

   The conclusion is that in order to obtain a unitary evolution
of states, we  are {\it required} to sum over 4-manifolds of both
Lorentzian and Euclidean signature, and in general over manifolds which
may be Lorentzian in some regions, and Euclidean in others.  This
raises the obvious question of why spacetime seems to have Lorentzian
signature, rather than Euclidean or mixed signature.

   The question "why is spacetime Lorentzian?" can be raised already
at the level of classical general relativity.  Einstein's equations
themselves do not specify a choice of metric signature; there are
Lorentzian solutions to these equations, and there are Riemannian
solutions.  Recently, solutions to the Einstein equations in which
part of the manifold is Riemannian (Euclidean signature) and the
rest is Lorentzian have been studied \cite{Ellis}; it is conceivable
that solutions of this kind are relevant to the very early Universe.
In any case, the signature of a manifold solving the Einstein
equations is determined in general from initial conditions $\{g_{ij},
p^{ij}\}$ satisfying the appropriate constraints.  A given initial
3-manifold may trace out either a Lorentzian or Riemannian 4-manifold,
depending on the initial choice of conjugate momenta.

    The dependence of lapse on initial conditions applies also to the
quantum theory.  The general solution of the evolution equation
\rf{Seq} for the "relativistic particle" example, with $\A$ given
in eq. \rf{relpart}, is
\beq
       \psi(x^\m,\t) = \int d^4p \; f(p)
\exp\left[-{i\over \hbar}\left({(D+1)
\over 2m^2}p^2\t+p_\m x^\m\right)\right]
\eeq
It is easy to see that
\beq
       <x^\m> = <x^\m>_0 + {(D+1)\over m^2}<p^\m>  \t
\eeq
(recall that $\t$ has units of action).  So long as $f(p)=0$ for
$p^2>0$, the expectation value of position follows a timelike path.
We would expect the same situation in quantum gravity, for the same
reason, namely, the Ehrenfest principle.  If the initial "wavefunction of the
Universe" $\Psi[q^a(x),\t_0]$ has expectation values which are peaked
around some (equivalence class of) configurations and momenta $\{q,p\}_0$,
then the wavefunction tends to remain peaked in the neighborhood
of a classical manifold which solves the Einstein equations for this
initial data.  Thus, despite the fact that the path integral sums over
Lorentzian and Euclidean manifolds, the probability density can still
be sharply peaked at one or the other signature.

   Obviously these remarks do not answer the question "why is spacetime
Lorentzian?", but only replace it with another question about initial
conditions.  For an attempt to explain the
preference for Lorentzian signature (in the context of
non-time-parametrized theories) from an analysis of an effective
"signature potential", see ref. \cite{Us2}.

\section{Quantum Theory in Curved Spacetime}

    Associated with the problem of time in quantum gravity is
a "problem of state."  Let us return, for a moment, to the
standard formulation of canonical quantum gravity, which in
our language is a restriction to a single value of $\E$, and
let $H$ be the Wheeler-DeWitt Hamiltonian.  Suppose a physical
state $\Psi$ is an eigenstate of an observable $Q$; this means
that $Q\Psi$ must also be a physical state.  But then
\bea
        H(Q\Psi) = [H,Q] \Psi = 0
\eea
which is not true, in general, unless $[H,Q]$ vanishes weakly.
It is then problematic to construct physical states which are
approximate eigenstates of, e.g., 3-geometry, or the position of
the hands of a clock.

   In this section we show how to construct physical states
which are sharply peaked around a given 3-geometry and extrinsic
curvature.  Treating the metric degrees of freedom semiclassically,
the dynamics of the other degrees of freedom approximates the standard
quantum theory on a curved background. Of course, the WKB treatment
can be extended to any other degrees of freedom (such as the hands
of a clock) which behave more or less classically.

    We recall that our path integral leads, in the end, to the
following solution for the evolution of physical states:
\beq
   \Psi[q,\t] = \sum_{\E ,\a} c[\E , \a] \Phi_{\E,\a}[q] e^{i\E \t/\hbar}
\eeq
where
\beq
        \A \Phi_{\E,\a} = -\E \Phi_{\E,\a}
\label{AE3}
\eeq
and where the subscript $\a$ is meant to distinguish between different
solutions of \rf{AE3}.  As discussed above, eq. \rf{AE3} is a
one-parameter $(\E)$  class of Wheeler-DeWitt equations
\beq
   \left[ -{\hbar^2 \over \E}\k^2 G^{ab} {\d^2 \over \d q^a \d q^b}
           + \rg U \right] \Phi_{\E,\a} = 0
\label{WD1}
\eeq
each of which can be treated by WKB methods.  To get the
quantum-theory-in-curved-spacetime limit, we follow the approach of Banks
\cite{Banks},
treating the metric semiclassically, and expanding in powers of $\k^2$
(back-reaction of matter on metric will be ignored; it can
presumably be dealt with following the approach of ref. \cite{Brout}).
Thus, write \rf{AE3} in the form
\beq
    \left\{ - \left[ {\hbar^2 \over \E}\k^2 G_{ijkl} {\d^2 \over
\d g_{ij} \d g_{kl} } + {1 \over \k^2}\rg R \right] + \H^\E_m \right\}
   \Phi_{\E,\a} = 0
\eeq
where $\H^\E_m$ is the Hamiltonian density for the non-gravitational fields,
denoted $\p$.  We then make the WKB ansatz
\beq
      \Phi_{\E,\gb}= \exp\left[ i\rE S[g,\gb]/\k^2 \hbar \right]
                        \rho^\E_{VV}[g] \psi^\E_m[\p]
\eeq
where $S[g,\gb]$ is Hamilton's principal function (the action of a
4-manifold solving the Einstein equations, bounded by the three
manifolds with metric $\gb_{ij}$ and $g_{ij}$); it satisfies the
Einstein-Hamilton-Jacobi equation in both arguments
\bea
      G_{ijkl}{\d S \over \d g_{ij}}{\d S \over \d g_{kl}}
                  - \rg R &=& 0
\non \\
      G_{ijkl}[\gb]{\d S \over \d \gb_{ij}}{\d S \over \d \gb_{kl}}
                  - \sqrt{\gb} R[\gb] &=& 0
\eea
The functional $\rho^\E_{VV}[g]$ is the Van-Vleck determinant, while
$\psi_m^\E$ is a solution of the Tomonaga-Schwinger equation
for quantum theory on a curved spacetime background
\beq
      i \hbar_{eff} {\d \psi^\E_m \over \d T(x;g,\gb)}
                = \H^\E_m \psi^\E_m
\label{TS}
\eeq
where
\beq
        \hbar_{eff} = {\hbar \over \rE}
\eeq
is the effective value of Planck's constant,
and $T(x;g,\gb)$ is a functional of the background spacetime defined by
\beq
 {\d \over \d T} = 2 G_{ijkl} {\d S \over \d g_{ij}}{\d \over \d g_{kl}}
\eeq
Up to this point, we have simply repeated the analysis of ref. \cite{Banks}

\bigskip

   However, the semiclassical approach to
recovering ordinary quantum field theory,
as outlined above, is subject to the following objection:  Although the part
of the wavefunction involving the non-gravitational fields obeys a
Tomonaga-Schwinger equation, the metric $g_{ij}$, on which the
"many-fingered" time parameter $T(x;g,\gb)$ depends, is still a
dynamical degree of freedom, and {\bf there is
no physical state satisfying the Wheeler-DeWitt equation which has a
probability distribution peaked at a $\underline{\mbox{\bf particular}}$
3-geometry $g_{ij}$}, i.e
the wavefunction is not peaked on any particular time-slice of a 4-manifold.
In fact, the squared-modulus of the leading term in the WKB approach, i.e.
\beq
\left| \exp\left[ i\rE S[g,\gb]/\k^2 \hbar \right]\right|^2 = 1
\eeq
has no dependence on $g_{ij}$ at all.  The best one can do in the standard
formulation (that is, using only a single value of $\E$)
is to superimpose WKB solutions
\bea
   \Phi_{\E,F}[g] &=& \int D\gb_{ij} F[\gb_{ij}]   \Phi_{\E,\gb}
\non \\
   &=& \int D\gb_{ij} f[\gb_{ij}] \exp\left[ i\{ \rE S[g,\gb]/\k^2
             + \th[\gb] \}/\hbar \} \right] \rho^\E_{VV} \psi^{\E}_m
\eea
where $f[\gb]$ is a real functional peaked (modulo diffeomorphisms)
at a particular 3-geometry $g_{0 ij}$, and we define
\beq
    p_0^{ij} \equiv \left({\d \th \over \d \gb_{ij}}\right)_{|\gb = g_0}
\label{p0}
\eeq
where $\th[\gb]$ is the phase of the smearing functional $F[\gb]$.
As shown many years ago by Gerlach \cite{Gerlach}, this
superposition is still not peaked at any one 3-geometry, but rather on
all three-geometries which are spacelike slices of a certain 4-manifold,
satisfying Einstein's equations with initial data $\{g_{0 ij},p_0^{ij}\}$.
Thus there is no physical state, and no subspace of physical states, which
would correspond to an eigenstate of a non-stationary observable (such as the
three-geometry, or the fields on a given three-geometry).

   It is at this point that we make use of the freedom, inherent in our
formulation, to superimpose states of different $\E$, and write
\bea
   \Psi[g_{ij},\p,\t] &=& \int d\E D\gb \; F[\gb,\E]
        \exp\left[ {i\over \hbar}\{ \E \t + \rE S[g,\gb]/\k^2 \} \right]
             \rho^\E_{VV} \psi^{\E}_m[\p,T(x;g,\gb)]
\non \\
         &\approx& \psi^{\E_0}_m[\p,T(x;g,g_0)]
                   \int d\E D\gb \; F[\gb,\E]
  \exp\left[ {i\over \hbar}\{\E \t + \rE S[g,\gb]/\k^2 \} \right]
              \rho^\E_{VV}
\non \\
         &=& \psi^{\E_0}_m[\p,T(x;g,g_0)] \Phi_F^{(g_0,p_0)}[g_{ij},\t]
\eea
where it is assumed that $F[\gb,\E]$ is sharply peaked around
$\E=\E_0,~\gb_{ij}=g_{0ij}$, and $p_0^{ij}$ is defined as in eq.
\rf{p0} above.
The $\psi_m$ factor can be pulled outside the integral, on the grounds
that its variation with $\E$ and $\gb$ is much less than that of the
smearing function $F[\gb,\E]$, and the $\exp[i\rE S/\k^2 \hbar]$ factor.
Now consider the leading WKB term
\beq
      \Phi_F^{(g_0,p_0)}[g_{ij},\t] = \int d\E D\gb_{ij} \;
          f[\E,\gb] \exp\left[{i\over\hbar} \left\{
    \E \t + \rE S[g,\gb]/\k^2 + \th[\gb] \right\} \right]\rho^\E_{VV}
\eeq
This wavefunction will be peaked at configurations $g_{ij}$ where the
phase in the integrand is stationary, with respect to small variations
in $\gb_{ij}$ and $\E$ around $g_{0ij}$ and $\E_0$, respectively.
In other words, the wavefunction is peaked at metrics $g_{ij}$,
at time $\t$, such that
\bea
         \t &=& - {1 \over 2\k^2 \rE_0} S[g,g_0]
\non \\
         p_0^{ij} &=& - {\rE_0\over \k^2}
    \left( {\d S[g,\gb] \over \d \gb_{ij} }\right)_{|\gb = g_0}
\label{stat_phase}
\eea
The second of these two equations is satisfied by the metric $g_{ij}$
of any time-slice of a 4-manifold, satisfying Einstein's equations with
initial data $\{g_{0ij},p_0^{ij}\}$.  The first equation requires that
the action of the 4-manifold between the initial slice $g_{0ij}$ and the given
slice $g_{ij}$ is proportional to the time-parameter $\t$.  Now consider
a foliation of the given 4-manifold parametrized by some variable $x_0$,
with $g_{ij}=g_{0ij}$ at $x_0=0$.  Hamilton's principal function
$S[g,g_0]$ is monotonic in $x_0$, which means that $S[g,g_0]=0$ only
for $g_{ij} = g_{0ij}$.
It follows that, at $\t=0$, eq. \rf{stat_phase} gives us
\beq
       0 = S[g,g_0] ~~ \Longrightarrow ~~ g_{ij} = g_{0ij} ~~~
          \mbox{(modulo diffeomorphisms)}
\eeq
As a consequence, at $\t=0$, the
wavefunction $\Phi_F^{(g_0,p_0)}[g,\t=0]$ is peaked at $g_{ij}=g_{0ij}$
(modulo diffeomorphisms).
Thus, from the definition of the many-fingered time variable,
where $T(x;g_0,g_0)=0$,
\bea
     \Psi[g_{ij},\phi,\t=0] &=& \Phi_F^{(g_0,p_0)}[g,\t=0]
                \times \psi_m^{\E_0}[\p,T(x;g,g_0)]
\non \\
             &\approx& \Phi_F^{(g_0,p_0)}[g,\t=0]
                \times \psi_m^{\E_0}[\p,T=0]
\label{pstates}
\eea

\bigskip

    The importance of eq. \rf{pstates} is that there exists, in our
formulation, a class of states where the metric (and extrinsic curvature)
is sharply peaked around a given geometry $g_{0ij}$ (and $p_0^{ij}$),
and where the state factorizes into a wavefunction ($\Phi_F$) suppressing
fluctuations away from the given 3-geometry, and a wavefunction ($\psi_m$)
describing the state of the non-gravitational fields on that 3-geometry.
Such states can be fairly described as eigenstates of non-stationary
observables; these eigenstates are impossible to construct, as physical states,
in the standard formulation of canonical quantum gravity.

\bigskip

    Finally, we consider transition probabilities.  Take an initial
state of the form
\beq
    \Psi_{in}[g_{ij},\p] = \Phi_F^{(g_0,p_0)}[g,0] \times \psi_m^{\E_0}[\p,0]
\eeq
and a final state of similar form
\beq
 \Psi_{f}[g_{ij},\p] = \Phi_{F'}^{(g'_0,p'_0)}[g,0] \times
                     {\psi'}_m^{\E_0}[\p,0]
\eeq
where the smearing function $F'$ is peaked around some time slice $(g'_0,p'_0)$
of the classical 4-geometry specified by the initial data $(g_0,p_0)$.
The transition probability for $\Psi_{in} \ra \Psi_f$ after a time $\t$ is
given
by the factorized expression
\bea
     P_{in \ra f}(\t) &=& |<\Psi_f|e^{-i\A \t/\hbar}|\Psi_{in}>|^2
\non \\
          &=& |<\Psi_f|\Psi_{in}(\t)>|^2
\non \\
          &=& |<\Phi_{F'}^{(g'_0,p'_0)}[g,0]|\Phi_F^{(g_0,p_0)}[g,\t]>|^2
\non \\
  &\times&  |<{\psi'}_m^{\E_0}[\p,0]|\psi_m^{\E_0}[\p,T(x;g'_0,g_0)]>|^2
\eea
The first of these factors
\beq
      |<\Phi_{F'}^{(g'_0,p'_0)}[g,0]|\Phi_F^{(g_0,p_0)}[g,\t]>|^2
\eeq
gives the probability, after a time $\t$,
to be on the time-slice described by $(g'_0,p'_0)$ (up to a
certain uncertainty, specified by the smearing function $F'$).
The second factor
\beq
     |<{\psi'}_m^{\E_0}[\p,0]|\psi_m^{\E_0}[\p,T(x;g'_0,g_0)]>|^2
\eeq
is the quantum-field-theory-in-curved-spacetime
result; it gives the probability for a transition from an initial
state $\psi_m$ of quantum fields on the time-slice $g_0$, to the
state $\psi'_m$ on the later time-slice $g'_0$.  Both the initial and
final 3-manifolds are time-slices of the same 4-manifold specified
by the initial data $\{g_0,p_0\}$, and the state $\psi_m[\p,T]$ evolves
according to the Tomonaga-Schwinger equation \rf{TS}.

   In this way, we see how approximate eigenstates of geometry and
extrinsic curvature may be constructed, and how the standard formalism
of quantum field theory in curved spacetime emerges.  We will not attempt
to go further and discuss the problem of measurement in this context,
apart from noting that any of the standard "realistic" approaches that
have been applied to non-relativistic quantum mechanics, e.g.
many-universes, decoherence, or Bohm's theory, can be applied in our
formulation as well.

\section{Inclusion of Fermions}

   We have so far assumed that the canonical momenta $p_a$ appear quadratically
in the Hamiltonian, with indices contracted by the supermetric.  The
Hamiltonian of a set of Dirac fields, on the other hand, is linear in the
fermionic momenta, and it is not immediately obvious how such fields are
incorporated into our approach.

   In our previous work \cite{Us} we found two independent methods for
determining the $\A$ operator.  The "undetermined constant" method
was based on the trivial observation that the actions $S$ and
$S'=\mbox{const.}\times S$ are equivalent at the classical level; this leads
to the fact that the ratio of the kinetic and potential terms of the
Hamiltonian (which is the $\A$ functional), is indeterminate at the classical
level.  The second method, leading to the same quantum theory, is the
transfer matrix method, whose "real-time" or Feynman version was
presented in the preceding sections.  We will now apply both methods to
obtain the $\A$ operator for gravity coupled to a Dirac field.

   The action for the Einstein-Dirac system is expressed in terms
of the fermion field $\psi$ and tetrad field $e^a_{~\m}$ as
\beq
     S_{ED} = \int d^4x \; \mbox{det}(e) \left[ ^4R + i\overline{\psi}
(e_a^{~\m} \gamma^a D_\m - m)\psi \right]
\eeq
where $D_\m$ is the usual covariant spinor derivative.
The extension of the canonical ADM formalism to this system was
worked out in ref. \cite{Nelson}, for the "time-gauge"
\beq
          e^0_{~i} = 0 ~~~~~(i=1,2,3)
\label{tg}
\eeq
In this gauge, the Einstein-Dirac action expressed in terms of
canonical momenta has the form
\beq
  S = \int d^4x \left[p_a \dot{q}^a + \pp \dot{\psi} -
(N\H + N_i \H^i + \e^{ij}M_{ij}) \right]
\eeq
\noindent where the $q^a$ are the triad fields $e^c_k(x)$, and
\bea
   \H &=& \k^2 G^{ab}p_a p_b + \rg U + \H_\psi
\non \\
\H_\psi &=& \pp K \psi = \pp \gamma^0
[e^i_a \gamma^a D_i - m] \psi
\eea
The first-class constraints are
\beq
       \H = \H^i = M_{ij} = 0
\eeq
where the supermomenta $\H^i$ and the generators of local frame rotations
$M_{ij}$ are linear in the momenta.
In addition there are 2nd-class constraints, some of which are associated
with the time-gauge condition \rf{tg}, and also which relate $\pp$ to
$\overline{\psi}$:\footnote{We take right derivatives with respect to $\psi$.}
\beq
         \pp = i\rg \overline{\psi} \gamma^0
\label{2con}
\eeq
The 2nd-class constraints are handled, according to the
Dirac procedure, by replacing Poisson brackets by Dirac brackets.
The explicit form of all constraints in terms of the canonical
variables, and other details, may be found in ref. \cite{Nelson}.

   Now consider an alternative action $S'_{ED}$ which differs from
$S_{ED}$ only by a multiplicative constant
\beq
           S'_{ED} = \sqrt{\E} S_{ED}
\eeq
Obviously, the equations of motion derived from $S'_{ED}$ are
identical to those derived from $S_{ED}$.  The constant
$\E$ is therefore irrelevant at the classical level.  In
going to the canonical formulation, however, we find that
\bea
  S' &=& \int d^4x \left[p_a \dot{q}^a + \pp \dot{\psi} -
(N\H^\E + N_i \H^i + \e^{ij}M_{ij} \right]
\non \\
  \H^\E &=& {1\over \ER}\k^2 G^{ab}p_a p_b + \H_\psi + \ER \rg U
\non
\eea
with 2nd class constraints enforcing
\beq
         \pp = i \sqrt{\E} \rg \overline{\psi} \gamma^0
\eeq
Define
\beq
     H^{\E} \equiv \int d^3x \;(N\H^\E + N_i \H^i + \e^{ij}M_{ij})
\eeq
and consider a field configuration $\{e^a_i(x,t),\psi(x,t),
\overline{\psi}(x,t)\}$ which
solves the Hamiltonian equations of motion derived from $H^\E$,
for some given value of $\E$.  Then it is clear that this
configuration is a solution for any other value of $\E$, since
the classical orbits in configuration space (i.e. solutions of the
Euler-Lagrange equations) are independent of $\E$.  In general then,
the Dirac bracket equation of motion
\footnote{Of course, $\{F,H^\E \}_D \approx \{F,H^\E \}$, since
$\H^\E = 0$ is a first-class constraint.}
\beq
      \partial_t F = \{F,H^\E \}_D
\eeq
supplemented by the first class constraints
\bea
      H^\E &=& 0  ~~~~ \mbox{for any} ~N,N_i,\e^{ij}
\non \\
    \Longrightarrow \H^\E &=& \H^i = M_{ij} = 0
\label{fclass}
\eea
generates a set of orbits in configuration space which is
independent of $\E$.  In this sense $\E$ is "classically
irrelevant."

  Now observe that the constraint $H^\E = 0$ can be written
\beq
       \A = -\E  ~~~~ \mbox{for any} ~N,N_i,\e^{ij}
\label{A=E}
\eeq
where $\A$ is defined implicitly by
\beq
 \A = \int d^3x \left\{ {N(\k^2 G^{ab}p_a p_b + \sqrt{-\A} \H_\psi )
\over \int d^3x' \; \rg NU}  + {1\over m_p}(N_i \H^i + \e^{ij}M_{ij})
\right\}
\label{AEpsi}
\eeq
and where $m_p$ is an arbitrary parameter with dimensions of mass.
 From this definition, it is straightforward to show that,
for any functional $F=F[q,p,\psi,\pp]$, the Poisson bracket with
$\A$ is related to the corresponding Poisson bracket with $H^\E$
via
\bea
  m_p \{F,\A\} &=& \int d^3x \biggl\{ \left[1+ {\int d^3x_1 N \H_\psi
\over 2 \ER \int d^3x_2 N\rg U } \right]^{-1}
{\sqrt{\E}m_p N\over \int d^3x_3
N \rg U} \{F,\H^\E\}
\non \\
   & & +  N_i \{F,\H^i\} + \e^{ij} \{F,M_{ij}\}
\biggr\}
\non \\
      &=&\{F,H^\E\}^{N\ra \N}
\label{Poissons}
\eea
where
\beq
    \N(N) \equiv \left[1+ {\int d^3x_1 N \H_\psi
\over 2 \ER \int d^3x_2 N\rg U } \right]^{-1} {\sqrt{\E}
m_p N\over \int d^3x_3
N \rg U}
\eeq
Eq. \rf{Poissons} is derived by simply carrying out the functional
derivatives contained in
the Poisson brackets shown, and applying the constraint
\rf{A=E}.  Then, since the Dirac bracket $\{F,\A\}_D$ is linear in
Poisson brackets $\{..,\A\}$, eq. \rf{Poissons} implies
\beq
      m_p \{F,\A \}_D = \{F,H^\E\}_D^{N\ra \N}
\eeq
Defining
$\t = m_p t$,
this demonstrates the equivalence of
\beq
   \partial_t O = \{O,H^\E\}_D ~~~~ \Leftrightarrow ~~~~
\partial_\t O = \{O,\A \}_D
\eeq
up to a time-reparametrization, expressed by $N\ra \N$.  Note that
$N(x)$ and $const. \times N(x)$ have the same $\N$.

   We now quantize by replacing Dirac brackets with commutators (in the
case of bosonic fields), and anticommutators (in the case of fermionic
fields).  Time evolution of states is given by the
Schr\"odinger equation
\beq
  i\hbar \partial_\t \Psi[q,\tilde{\psi},\t] = \A \Psi[q,\tilde{\psi},\t]
\eeq
with the general solution
\beq
      \Psi[q,\tilde{\psi},\t] = \sum_\E a_\E \Phi_\E [q,\tilde{\psi}]
                                      e^{i\E \t /\hbar}
\eeq
where
\beq
       \tilde{\psi} \equiv g^{\oq} \psi
\eeq
and $\Phi_\E$ satisfies a
Wheeler-DeWitt equation
\beq
   {1 \over \sqrt{\E}} \H^\E \Phi_\E =
   \left[ -{\hbar^2\over \E} \k^2 "G^{ab} {\d^2 \over \d q^a \d q^b}"
+ i{\hbar \over \sqrt{\E}} {\d \over \d \tilde{\psi}} K \tilde{\psi}
+ \rg U \right] \Phi_\E = 0
\label{WDfermion}
\eeq
where the operator-ordering remains to be specified.  Note that, as
in the purely bosonic case, the classically irrelevant constant $\E$
can be absorbed into a redefinition of $\hbar$
\beq
        \hbar_{eff} = {\hbar \over \ER}
\eeq
This concludes the first, "undetermined constant" method for
finding the $\A$ operator.

   Next we apply the path-integral approach, following as
closely as possible the procedure of the previous section
for the purely bosonic case.
Since the generalized BSW action for gravity + fermions will
contain a factor of $\H_{\psi}$ inside the square-root,
our strategy will be to expand the path-integrand
in powers of $\H_{\psi}$, and evaluate the relevant expressions
to some finite order (first order, in this article).  These expressions
can then be compared, order by order in $\H_{\psi}$, with results of the
"undetermined constant"  method above.

   We again set $N_i=0$, and also $\e^{ij}=0$, which is to be
compensated by imposing the corresponding physical state
constraints
\beq
      \H^i \Psi = 0 ~~~~~~ M_{ij} \Psi = 0
\eeq
Solving for the bosonic momenta in terms of the
time-derivatives
\beq
     p_a = {1 \over 2\k^2 N} G_{ab}\pa_t q^b
\eeq
inserting into the Hamiltonian constraint
\beq
{1 \over 4 \k^2 N^2} G_{ab} \pa_t q^a \pa_t q^b + \rg U + \H_\psi = 0
\eeq
and solving for the lapse function
\beq
   N = {1 \over 2\k} \left[ - {1 \over (\rg U + \H_\psi)} G_{ab} \pa q^a
\pa q^b \right]^{1/2}
\eeq
we arrive at a square-root action
\beq
   S = \int d^4x \left\{ i\rg \ops \gamma^0 \pa_t \psi - {1\over \k}
\sqrt{-(\rg U + \H_\psi)G_{ab} \pa_t q^a \pa_t q^b} \right\}
\eeq
The corresponding path-integral is
\bea
   \Psi[q',\tp',\t+\D \t] &=& \int Dq D\ops D\psi \; \m(q,\psi,\ops)
\exp\left[c_0\int d^4x \left\{ i\rg \ops \gamma^0 \pa_t \psi \right. \right.
\non \\
& & - {1\over \k} \left. \left.
\sqrt{-(\rg U + \H_\psi)G_{ab} \pa_t q^a \pa_t q^b} \right\} \right]
\Psi[q,\tp,\t]
\eea
where
\beq
\H_\psi = i\rg \ops \gamma^0 K \psi
\eeq
and $c_0$ represents the $\e \ra 0$ continuum limit of the
regularization-dependent constant $c_\e \propto 1/\sqrt{-i\e \hbar}$.
Now expand the exponential to first order in $\H_\psi$
\bea
   \lefteqn{ \Psi[q',\tp',\t+\D \t] }
\non \\
 &=& \int Dq D\ops D\psi \; \m(q,\psi,\ops)
\non \\
& & \times \left[1 - {c_0 \over \k} \int d^3x d\t \;
{\H_\psi \over 2\rg U} \sqrt{- \rg U G_{ab} \pa_\t q^a \pa_\t q^b} \right]
\non \\
& & \times \exp\left[c_0\left (S_0 +\int d^3x d\t \; i\rg \ops \gamma^0
\pa_\t \psi \right )\right] \Psi[q,\tp,\t]
\eea
where $S_0$ is the bosonic action
\beq
  S_0 = -{1\over \k}\int d^3x d\t \sqrt{-\rg U G_{ab} \pa_\t q^a \pa_\t q^b}
\eeq

   We now regularize the path-integral according to the lattice prescription
\rf{discrete}.  The bosonic part of this path-integral, based on the
action $S_0$, leads to the operator evolution
\beq
     i\hbar \pa_\t \Psi  = A \Psi
\eeq
where
\bea
      A &=& {(D+1)\over 2} \sum_n
        {\rg \over U} \k^2 G^{ab} {1\over v_n^2} (-\hbar^2)
        {\pa^2 \over \pa q^a(n) \pa q^b(n)}
\non \\
        &=& {(D+1)\over 2} \sum_n {\rg \over U} \k^2
            {1\over v_n^2} G^{ab} p_a^L(n) p_b^L(n)
\eea
and we have defined
\beq
       p_a^L(n) \equiv -i\hbar {\pa \over \pa q^a(n)}
\label{star}
\eeq
Then to zeroth-order in $\H_\psi$, we can identify the
derivatives $\pa_\t q$ in the term
proportional to $\H_\psi$ as proportional to the bosonic momentum
operators, according to
\bea
       \pa_\t q^a(n) &=& {\pa A \over \pa p_a^L(n)}
\non \\
           &=& (D+1)\k^2 {\rg \over U v_n^2}  G^{ab}p_b^L(n)
\non \\
           &=& {(D+1) \k^2 \over U v_n} G^{ab} p_b(n)
\eea
where we have introduced
\beq
       p_a(n) \equiv {\rg \over v_n} p_a^L(n)
\eeq
so that, as operators, using eq. \rf{star},
\beq
       p_a(n) \ra -i\hbar \left({\d \over \d q^a(n)}\right)_R
\eeq
The regularized path-integral, to first-order in $\H_\psi$ is
now
\bea
   \Psi[q',\tp',\t+\e] &=& \int Dq D\ops D\psi \; \m_\e(q)
(1 - \e W) \exp \left[-c_\e \sum_n v_n \{ i\ops \gamma^0 \D \psi \right.
\non \\
   & & \left. +{1\over \k} \sqrt{-\G_{ab} \D q^a \D q^b} \} \right]
\Psi[q'+\D q, \tp'+\D \tp ,\t]
\eea
where
\beq
    W = c_\e {(D+1)\over 2} \sum_n {1 \over \rg U}
\sqrt{-\k^2 G^{ab} p_a p_b \over \rg U}[i\rg \ops \gamma^0 K
(\psi' + \D \psi)]
\eeq
Carrying out the integrals over $\D q$, $\D \psi$, $\ops$, we find
\bea
\Psi_{\t+\e} &=& \left[1 + \e {(D+1) \over 2} \sum_n \left\{
{1\over i\hbar}{1\over \rg U} \k^2 G^{ab}p_a p_b \right. \right.
\non \\
& & \left. \left.  + {1\over \rg U} \sqrt{-\k^2 G^{ab}p_a p_b
\over \rg U} \left({\d \over \d \tp}\right)_R K \tp
\right\} \right] \Psi_\t
\eea
where the square-root operator is defined via spectral analysis.

   Once again, the requirement that the state evolution is independent,
in the continuum limit, of the choice of $\{ v_n \}$, implies that
the term in braces is the same in each cell (simplex) $n$.  Therefore, in
the continuum limit,
\beq
   i\hbar \pa_\t \Psi =  {1 \over \rg U}\left[
\k^2 G^{ab}p_a p_b + \sqrt{-\k^2 G^{ab}p_a p_b
\over \rg U} i\hbar {\d \over \d \tp} K \tp \right] \Psi
    ~~~~~~~ \mbox{(all $x$)}
\eeq
where the divergent factor $N_p (D+1)/2$ has been absorbed into a rescaling
of $\t$.  For stationary states $\Phi_\E$ we have
\beq
     \left[ \k^2 G^{ab}p_a p_b + \sqrt{-\k^2 G^{ab}p_a p_b
\over \rg U}i \hbar {\d \over \d \tp} K \tp + \E \rg U \right] \Phi_\E = 0
\eeq
Then, since
\beq
\sqrt{-\k^2 G^{ab}p_a p_b \over \rg U} \Phi_\E = \sqrt{\E}\Phi_\E
  ~~+~~ \mbox{order $\H_\psi$ corrections}
\eeq
it follows that, up to first order in $\H_\psi$, we recover
the same stationary state equation
\beq
   \left[ -{\hbar^2\over \E} \k^2 "G^{ab} {\d^2 \over \d q^a \d q^b}"
+ i{\hbar \over \sqrt{\E}} {\d \over \d \tilde{\psi}} K \tilde{\psi}
+\rg U \right] \Phi_\E = 0
\eeq
that was obtained (eq. \rf{WDfermion}) from the
"undetermined constant" approach.
As in the purely bosonic case, the imposition of $\H^\E \Phi_\E = 0$
for all $\E$ at every point $x$ is consistent with, and in fact implies
(c.f. ref. \cite{Moncrief}), the other required first-class constraints
$\H^i \Psi = M_{ij} \Psi = 0$, up to the usual operator-ordering issues.

\section{Conclusions}

   For ordinary quantum theories without time-parametrization,
the regularized path integral is expressed as a product of integrals,
each of which evolves the state function unitarily over a
very small time interval.  In this article we have examined whether
such a construction can be applied to theories with square-root,
time-reparametrization invariant actions.  Our result is that
unitarity requires: i) an unconventional phase in the path-integrand;
and ii) summation over configurations of both real and imaginary
proper-time lapse.  In the case of quantum gravity, the second requirement
means that path-integration must run over manifolds of
Lorentzian, Euclidean and, in general, mixed signature.  We have also
shown how the formalism extends to fermionic actions.

   Unitarity, of course, refers to evolution in a certain time
parameter.  In our formulation, the time parameter is simply a
measure of the number of integrations in the (regulated) path-integral,
evolving an initial state to a later state.  This "quantum time" parameter
is neither a geometrical quantity (such as a proper time lapse), nor
a dynamical variable (such as the extrinsic curvature). It is, instead,
a parameter which is intrinsic to the path-integral measure.  The
connection to classical dynamics is established via an Ehrenfest principle.

   In the standard canonical formulation of quantum gravity, the physical
states are solutions of a Wheeler DeWitt equation $\H \Psi = 0$.
In contrast, an outcome
of our formulation is that physical states belong to a Hilbert space
which is spanned by the solutions of a {\it family} of Wheeler-DeWitt
equations $\H^\E \Phi_\E = 0$, which are distinguished by having
different effective values of Planck's constant
$\hbar_{eff}=\hbar / \sqrt{\E}$.
As discussed in section 4, a superposition of states with varying
$\E$ (or $\hbar_{eff}$) allows us to construct, at the semiclassical
level, physical states whose amplitudes are peaked at particular
3-geometries and extrinsic curvatures.  The width of the peak, in superspace,
is inversely proportional to the dispersion
$\D \E$.    Projection operators formed from
such states, and linear combinations of those projection operators, belong to
the physical observables of the theory.  It is worth noting that the stationary
states (i.e. solutions of a Wheeler-DeWitt equation with a fixed value
of $\E$) can never be peaked around any one 3-geometry.  At best, in
the WKB limit, a stationary state is peaked at every possible
spacelike slice of some 4-manifold satisfying the Einstein equations.

   If our view is correct, then the phenomenological value of
Planck's constant is the mean value of a dynamical quantity,
having a finite uncertainty of quantum origin.  How large this
uncertainty might be, and whether there could conceivably be
testable consequences, are interesting issues for further
investigation.

\vspace{33pt}

\noindent {\Large \bf Acknowledgements}{\vspace{11pt}}

   We would like to thank Maurizio Martellini for some very stimulating
discussions.  J.G. is also happy to acknowledge the hospitality
of the Lawrence Berkeley Laboratory, and the Niels Bohr Institute.
J.G.'s research is supported in part by  the U.S. Dept. of Energy, under
Grant No. DE-FG03-92ER40711.
A.C.'s research is supported by an EEC fellowship in the `Human Capital
and Mobility' program, under contract No. ERBCHBICT930313.

\end{document}